\documentclass[12pt]{article}
\usepackage[left=1in,right=1in,bottom=1in,top=1in]{geometry}

\usepackage{amsmath}
\usepackage{amssymb}
\usepackage{amsthm}
\usepackage{graphicx}
\graphicspath{{code/out/}}
\usepackage{natbib}
\usepackage{hyperref}
\hypersetup{colorlinks=true, citecolor=black, linkcolor=., urlcolor=cyan}
\usepackage[font = small]{caption}
\usepackage[shortlabels]{enumitem}
\usepackage{booktabs}
\usepackage{algorithm}
\usepackage{algpseudocode}
\algdef{SxnE}[FOR]{For}{EndFor}[1]{\algorithmicfor\ #1 }
\usepackage{tikz}
\usetikzlibrary{tikzmark}
\usepackage{array} 
\usepackage[normalem]{ulem}
\usepackage{appendix}
\usepackage{url}

\algnewcommand\algorithmicswitch{\textbf{switch}}
\algnewcommand\algorithmiccase{\textbf{case}}
\algnewcommand\algorithmicassert{\textbf{then }}
\algnewcommand\Assert[1]{\State \algorithmicassert #1}%
\algnewcommand\Assertif[2]{\State \textbf{if} #1 \algorithmicassert #2}
\algnewcommand\Assertelse[1]{\State \textbf{else} #1}
\algdef{SE}[SWITCH]{Switch}{EndSwitch}[1]{\algorithmicswitch\ #1}{\algorithmicend\ \algorithmicswitch}%
\algdef{SE}[CASE]{Case}{EndCase}[1]{\algorithmiccase\ #1}{\algorithmicend\ \algorithmiccase}%
\algtext*{EndSwitch}%
\algtext*{EndCase}%

\renewcommand{\H}{\mathbf{H}}
\providecommand{\I}{\mathbf{I}}

\providecommand{\Q}{\mathbf{Q}}

\providecommand{\W}{\mathbf{W}}
\providecommand{\X}{\mathbf{X}}

\renewcommand{\r}{\mathbf{r}}
\newcommand{\s}{\mathbf{s}}

\newcommand{\x}{\mathbf{x}}
\newcommand{\y}{\mathbf{y}}

\let\origc\c
\DeclareRobustCommand\c{\ifmmode\mathbf{c}\else\expandafter\origc\fi}
\let\origd\d
\DeclareRobustCommand\d{\ifmmode\mathbf{d}\else\expandafter\origd\fi}
\let\origu\u
\DeclareRobustCommand\u{\ifmmode\mathbf{u}\else\expandafter\origu\fi}
\let\origd\v
\DeclareRobustCommand\v{\ifmmode\mathbf{v}\else\expandafter\origv\fi}

\providecommand{\cC}{\mathcal{C}}

\providecommand{\bb}{\boldsymbol{\beta}}
\providecommand{\bh}{\widehat{\beta}}
\providecommand{\bbh}{\widehat{\boldsymbol{\beta}}}

\providecommand{\veps}{\varepsilon}

\providecommand{\bvep}{\boldsymbol{\varepsilon}}

\providecommand{\lam}{\lambda}

\providecommand{\cor}{\textrm{Cor}}

\providecommand{\Norm}{\textrm{N}}

\providecommand{\diag}{\textrm{diag}}
\providecommand{\Tr}{^{\scriptscriptstyle\top}}

\providecommand{\CV}{\textrm{CV}}

\usepackage[scr=boondox]{mathalpha}

\makeatletter
\@ifpackagelater{mathalpha}{2021/01/01}{%

}{%

}
\makeatother

\providecommand{\iid}{\overset{\text{iid}}{\sim}}

\providecommand{\sign}{\textrm{sign}}

\providecommand{\abs}[1]{\left\lvert#1\right\rvert}

\providecommand{\norm}[1]{\lVert#1\rVert}

\providecommand{\as}[1]{\begin{align*}#1\end{align*}}

\newlength{\li} 

\renewcommand{\abstract}[1]{
 \centerline{
 \begin{minipage}{0.7\linewidth}
 \hrule
 \vskip 0.1in
  \begin{center}
    {\bf Abstract}
  \end{center}
  #1
 \vskip 0.1in
 \hrule
 \end{minipage}}
 \vskip 0.3in}

\title{Alternative Likelihood Approximations for High-Dimensional Intervals for Lasso}
\author{
  Logan Harris\\Department of Biostatistics\\University of Iowa
  \and
  Patrick Breheny\\Department of Biostatistics\\University of Iowa
}
\date{\today}

\begin{document}

\maketitle

\abstract{
Classical frequentist approaches to inference for the lasso emphasize exact coverage for each feature, which requires debiasing and severs the connection between confidence intervals and the original lasso estimates. To address this, in earlier work we introduced the idea of average coverage, allowing for biased intervals that align with the lasso point estimates, and proposed the Relaxed Lasso Posterior (RL-P) intervals, which leverage the Bayesian interpretation of the lasso penalty as a Laplace prior together with a Normal likelihood conditional on the selected features. While RL-P achieves approximate average coverage, its intervals need not contain the lasso estimates. In this work, we propose alternative constructions based on different likelihood approximations to the full high-dimensional likelihood, yielding intervals that remain centered on the lasso estimates while still achieving average coverage. Our results continue to demonstrate that intentionally biased intervals provide a principled and practically useful framework for inference in high-dimensional regression.
}

\section{Introduction}

The lasso (least absolute shrinkage and selection operator) is a popular penalized regression method that applies the $L_1$ penalty, $\lambda\norm{\bb}_1$, to a loss function. In lasso-penalized linear regression \citep{Tibshirani1996}, the objective function is $$Q(\bb|\X,\y,\lambda) = \tfrac{1}{2n}\norm{\y - \X\bb}_2^2 + \lambda\norm{\bb}_1,$$ where $\y$ is a length $n$ vector of independent outcomes, $\X$ is an $n \times p$ matrix of features, $\bb$ is a length $p$ vector of regression coefficients, and $\lambda$ is a regularization parameter controlling the amount of penalization. The resulting fit is generally sparse, a hallmark of the $L_1$ penalty. However, because of complexities introduced by the $L_1$ penalty, methods for obtaining intervals for the lasso have lagged well behind the usage of the lasso itself. 

The classical frequentist definition of coverage insists that a confidence interval maintain nominal coverage for all possible values of the parameter. Attempting to satisfy this requirement requires some form of debiasing to be applied to the lasso in order to counteract shrinkage \citep{ZhangZhang2014, Javanmard2014, Lee2016}. The consequence of this debiasing is that the agreement between the intervals and the lasso point estimates is often very poor. To address this, \citet{AvgCov} proposed shifting focus from the coverage of individual parameters to the average coverage over the entire set of parameters in the model. Intervals providing such average coverage were termed high-dimensional intervals (HDIs) as they strike a middle ground between confidence and credible intervals. Building on that framework, the paper then constructed Relaxed Lasso Posterior (RL-P) intervals by exploiting the connection between the lasso penalty and a Laplace prior to construct distributions for $\beta_j$ conditional the selected model. These distributions are based on a mild extension of the likelihood used by the Relaxed Lasso, giving rise to the name, Relaxed Lasso Posterior.

The RL-P method performs well in a wide range of scenarios under the newly proposed definition of coverage. However, the Relaxed Lasso likelihood is not the only --- and not necessarily the most appropriate --- one with which to construct parameter intervals for the lasso. In particular, its posterior mode is not equal to the lasso point estimate. In severe cases, this inconsistency can lead to situations in which the interval does not contain its corresponding lasso point estimate.

In Section~\ref{sec:proj-int}, we propose two alternative interval construction methods. These methods use the same posterior construction as the RL-P intervals, but with different likelihoods. The construction guarantees that the resulting HDIs always contain the lasso estimates, as the lasso estimates are the modes of the distributions used to construct the intervals. In Section~\ref{sec:results}, we then compare these methods to the RL-P intervals. In Section~\ref{sec:biased-intervals}, we compare biased vs unbiased approaches to interval construction and show that debiasing does not necessarily lead to superior coverage in practice. Finally, in Section~\ref{Sec:Scheetz2006-2}, we compare the three average coverage methods proposed so far using an application to a data set on gene expression data in mammalian eyes.

\section{Projection Based High-Dimensional Intervals}
\label{sec:proj-int}

Viewed from a Bayesian perspective, the lasso penalty corresponds to a Laplace (double-exponential) prior on the regression coefficients. Combined with the normal likelihood from linear regression, this formulation provides a probabilistic framework that can be used to construct intervals. Because the prior is determined by the penalty once $\lam$ has been selected, improving upon RL-P intervals requires modifying the likelihood. Although the lasso objective also implies a likelihood, it is difficult to work with directly --- when constructing an interval for $\beta_j$, the remaining $p-1$ elements of $\bb$ act as nuisance parameters. The central question then is which aspects of the selected model should be conditioned on when approximating this likelihood for an individual $\beta_j$. If we further require that the lasso estimate represents the posterior mode, the likelihood's center is also fixed. What remains is to determine an appropriate variance in this likelihood, and in this section, we consider two approaches.

In Section~\ref{sec:pipep}, we propose a likelihood based on work in \cite{dai2019}. In both this method (PIPE-P) and RL-P, the variance is based on a full projection onto the active features. However, in penalized regression, the active features can also be seen as existing on a continuum between inactive and fully active, so in Section~\ref{Sec:LQAP} we propose a method that estimates the likelihood variance based on partial projections onto the active set.

Without loss of generality, we assume $\X$ is standardized so that $\x_j \Tr \x_j = n$ and $\y$ is centered, so no intercept is required.

\subsection{PIPE Posterior}
\label{sec:pipep}

The likelihood for $\beta_j$, obtained by conditioning on all other parameters treated as fixed at the lasso point estimates, $\bbh_{-j}$, is:
$$ L(\beta_j|\bbh_{-j}) \propto \exp(-\tfrac{n}{2\sigma^2}(\beta_{j} - z_{j})^2), $$

\noindent where $z_j = \x_j \Tr\r_{-j} / n$ is the correlation between feature $j$ and the partial residuals obtained by excluding feature $j$, $\r_j = \y - \X_{-j}\bbh_{-j}$, and $\sigma^2$ is the error variance. We refer to this likelihood as the full conditional likelihood. With the full conditional likelihood and the Laplace prior
\as{p(\bb) = \prod_{j = 1}^{p} \tfrac{\gamma}{2}\exp(-\gamma \abs{\beta_j})}

\noindent where $\gamma = n\lam / \sigma^2$, $\bbh(\lam)$ is the posterior mode for $\bb$. This satisfies the criteria of having a distribution which has its mode equal to the lasso point estimates. However, this approach treats the remaining parameters as fixed. In effect, it assumes the features are orthogonal and ignores the uncertainty they introduce for parameter $\beta_j$. Consequently, outside the orthogonal setting, intervals based on the resulting posterior fail to provide valid average coverage. 

To get around these issues, \cite{AvgCov} use a likelihood that accounts for the selected features. Let $\hat{S} = \lbrace k: \bh_k \neq  0 \rbrace$ denote the set of selected features and $\hat{S}_j$ denote the set of selected features that excludes feature $j$, so that $\hat{S}_j = \hat{S} \text{ if } j \notin \hat{S}$ and $\hat{S}_j = \hat{S} - \lbrace j \rbrace \text{ if } j \in \hat{S}$. Now, define $\Q_{\hat{S}_j} = \I - \X_{\hat{S}_j}(\X_{\hat{S}_j} \Tr \X_{\hat{S}_j})^{-1} \X_{\hat{S}_j} \Tr$, the projection matrix onto the features selected by the lasso excluding feature $j$. Then, the likelihood for $\beta_j$ conditional on the selected features is: 
\as{L(\beta_j|\hat{S}_j) \propto \exp\left(-\frac{\x_j \Tr \Q_{\hat{S}_j} \x_j}{2\sigma^2}(\beta_{j}^2 - \tilde{\beta}_j)^2\right),}

\noindent where $\tilde{\beta}_j = (\x_j \Tr \Q_{\hat{S}_j} \x_j)^{-1} \x_j \Tr \Q_{\hat{S}_j} \y$.  This can be seen as an extension of the Relaxed Lasso, and the resulting intervals were named Relaxed Lasso Posterior (RL-P) intervals. Intervals produced with this likelihood have approximately correct average coverage. However, the resulting posterior mode is different than the point estimate produced by the lasso fit.

Many debiasing approaches require refitting or solving additional systems for each coefficient, which is computationally intensive. PIPE \citep{dai2019} was introduced as a projection-based alternative to avoid repeated refitting while still yielding valid test statistics. By leveraging the KKT conditions of penalized regression, PIPE constructs an estimator for each coefficient that is asymptotically normal. Specifically, the PIPE estimator for $\beta_j$ is centered at $z_j$ and has a variance $\sigma^2/ (\x_j \Tr \Q_{\hat{S}_j} \x_j)$. Accordingly, it can be viewed as implying the following Normal likelihood:
\as{L(\beta_j|\hat{S}_j) \propto \exp\left(-\frac{\x_j \Tr \Q_{\hat{S}_j} \x_j}{2\sigma^2}(\beta_{j} - z_{j})^2\right).} 

\noindent This likelihood is a middle ground between the full conditional likelihood and the RL-P likelihood. It has the same mean as the full conditional likelihood and the same variance as the RL-P likelihood. As a result, posteriors built off of this PIPE based likelihood have the mode equal to the estimate from the original lasso fit but adjust for the uncertainty given the selected covariates. We call the intervals produced using this likelihood PIPE Posterior (PIPE-P) intervals.

With this likelihood, the resulting posterior can be shown to be
\as{
p(\beta_j | \hat{S}_j) &\propto
\begin{cases}
C_{-} \exp\bigl\{-\frac{\x_j \Tr \Q_{\hat{S}_j} \x_j}{2\sigma^2} (\beta_j - (z_j + \lambda))^2\bigr\}, \text{ if } \beta_j < 0, \\
C_{+} \exp\bigl\{-\frac{\x_j \Tr \Q_{\hat{S}_j} \x_j}{2\sigma^2} (\beta_j - (z_j - \lambda))^2\bigr\}, \text{ if } \beta_j \geq 0 \\
\end{cases}
}
where $C_{-} = \exp(z_j \lambda \x_j \Tr \Q_{\hat{S}_j} \x_j / \sigma^2)$ and $C_{+} = \exp(-z_j \lambda \x_j \Tr \Q_{\hat{S}_j} \x_j / \sigma^2)$.

The details can be found in \cite{AvgCov}, but this formulation is attractive because it allows efficient computation of quantiles -- although the distribution does not have a standard parametric form, the left and right components do, and this can be utilized to compute quantiles analytically.

Of course, the above formulation depends on a specific value of $\lam$. We recommend performing cross validation and selecting the $\lam$ which minimizes the cross validation error ($\lam_{\CV}$). Likewise, the resulting posterior also depends on an estimate for $\sigma^2$. Here, as in the original PIPE manuscript, we take the suggestion of \cite{Reid2016} and estimate $\sigma^2$ with $\hat{\sigma}^2 = ||\y - \X\bbh({\lambda_{\CV}})||_2^2 / (n - |\hat{S}_{\CV}|)$, where $\abs{\hat{S}_{\CV}} = \#\{ \bh_j({\lambda_{\CV}}) \neq 0 \}$.

Putting this all together we can arrive at the following implementation to construct HDIs with significance level $\alpha$:

\begin{algorithm}[!ht]
\caption{PIPE-P Intervals at a single CV-chosen \(\lambda\)}
\label{alg:singlelambda-ci}
\begin{algorithmic}[1]
\Require \(\X\in\mathbb{R}^{n\times p},\;\y\in\mathbb{R}^n,\;\alpha\in(0,1)\)
\Ensure \(\{\mathrm{CI}_j\}_{j=1}^p\)
  \State \(\lambda_{\mathrm{CV}}\gets\arg\min_{\lambda}\mathrm{CV\mbox{-}Error}(\lambda)\) 
    \Comment{10-fold CV on \((\X,\y)\)}
  \State \(\hat\beta\gets\mathrm{Lasso}(\X,\y;\lambda_{\mathrm{CV}})\)
  \For{\(j=1,\dots,p\)}
    \State \(\r_{-j}\gets \y - \X_{-j}\,\bbh_{-j}\)
    \State \(z_j\gets \tfrac{1}{n}\,\x_j^\top \r_{-j}\)
    \State \(\hat S_j\gets\{\,k\neq j : \hat\beta_k\neq0\}\)
    \State \(\Q_{\hat S_j}\gets I_n - \X_{\hat S_j}(\X_{\hat S_j}^\top \X_{\hat S_j})^{-1}\X_{\hat S_j}^\top\)
    \State \(\mathrm{CI}_j\gets \bigl[P^{-1}(\tfrac{\alpha}{2}\mid\hat S_j),\,P^{-1}(1-\tfrac{\alpha}{2}\mid\hat S_j)\bigr]\)
  \EndFor
  \State \Return \(\{\mathrm{CI}_j\}_{j=1}^p\)
\end{algorithmic}
\end{algorithm}

\subsection{Local Quadratic Approximation Posterior}
\label{Sec:LQAP}

PIPE-P is sensitive to feature inclusion: as additional features enter the selected model, the width of the resulting interval increases even if those features have estimates close to zero due to shrinkage. This is most pronounced in the near-saturated case, where $\x_j$ provides little information orthogonal to $\cC(\X_{\hat{S}_j})$. Feature correlation further exacerbates the problem: stronger correlations shrink the projection of $\x_j$ onto $\cC^\bot(\X_{\hat{S}_j})$, leaving little independent information which results in wide intervals with conservative coverage behavior.

This sensitivity arises because PIPE-P computes variance using a full projection onto the active features. Projecting $\x_j$ directly onto $\Q_{\hat S_j}$ discards all information not orthogonal to the selected features. On the other hand, a ``partial'', or regularized, projection would account for the fact that selected features are not fully active, but rather have been shrunk and only remove a portion of that information. PIPE-P calculates the variance based on a full projection but the mean based on a partially projection (i.e., uses the lasso estimates in computing the mean) whereas both the mean and variance of RL-P are based on a full projection. Here we consider an approach in which both the mean and variance are based on a partial projection.

To this end, we adopt the local quadratic approximation (LQA) of \cite{FanLi2001}. The goal here is to form a quadratic approximation for the lasso penalty and evaluate it at $\bh_k$ for $k \in \hat{S}_j$ and then use this to get a partial projection matrix onto $\cC^\bot(\X_{\hat{S}_j})$ in order to set the variance in the Normal likelihood. Because $\bh_k \neq 0$ for $k \in \hat{S}_j$, $\abs{\beta_k}$ is differentiable at $\bh_k$. Equivalently, this can be thought of as approximating the lasso penalty by the smallest quadratic majorant 
$$q(\beta_k) = \abs{\bh_k} + \text{sign}(\bh_k)(\beta_k - \bh_k) + c(\beta_k - \bh_k)^2$$

\noindent where $\beta_k \neq 0$ and $c > 0$ is chosen so that $q(\beta_k) \geq \abs{\beta_k}$ for all $\beta_k$. This occurs when $q(\beta_k)$ is also tangent to $\abs{\beta_k}$ at the mirror point $-\bh_k$, i.e. $q(-\bh_k) = \bh_k$. Solving this tangency condition yields
$$
\begin{aligned}
\abs{\bh_k} &= \abs{\bh_k} + \text{sign}(\bh_k)(-2\bh_k) + 4c\bh_k^2 \\
\Rightarrow 4c\bh_k^2 &= -\text{sign}(\bh_k)(-2\bh_k) \\
\Rightarrow c &= \frac{1}{2\abs{\bh_k}}.
\end{aligned}
$$

\noindent Thus, we have the following quadratic approximation for $\abs{\beta_k}$:
$$
\abs{\beta_k} \approx \abs{\bh_k} + \sign(\bh_k)(\beta_k - \bh_k) + \frac{1}{2}\frac{(\beta_k - \bh_k)^2}{\abs{\bh_k}}.
$$

\noindent Summing over $j \in \hat{S}_j$ we get:
$$
\begin{aligned}
\lam \sum_{k \in \hat{S}_j}\abs{\beta_k} &\approx \lam \s_{\hat{S}_j} \Tr \bb_{\hat{S}_j} + \tfrac{\lam}{2} (\bb_{\hat{S}_j} - \bbh_{\hat{S}_j}) \Tr \W^{-1} (\bb_{\hat{S}_j} - \bbh_{\hat{S}_j}) + C \\
&= \lam \s_{\hat{S}_j} \Tr \bb_{\hat{S}_j}  - \lam \bbh_{\hat{S}_j} \Tr\W^{-1} \bb_{\hat{S}_j} + \tfrac{\lam}{2} \bb_{\hat{S}_j} \Tr \W^{-1} \bb_{\hat{S}_j} + C^* \\
&= \tfrac{1}{2} \bb_{\hat{S}_j} \Tr \lam\W^{-1} \bb_{\hat{S}_j} + C^*, \\
\end{aligned}
$$

\noindent where $\W = \diag(|\bh_k|)_{k \in \hat{S}_j}$. This leads to the following loss surrogate:
$$
\min_{\beta_{\hat{S}_j}} \left[ \tfrac{1}{2n}\norm{\y - \X_{\hat{S}_j} \bb_{\hat{S}_j}}_2^2 + \tfrac{1}{2} \bb_{\hat{S}_j} \Tr \lam\W^{-1} \bb_{\hat{S}_j} \right].
$$

\noindent This form is the same as a Ridge regression with a weighted penalty, so as a function of $\y$ the fitted values can be written as:
$$
\hat{\y} = \X_{\hat{S}_j}(\tfrac{1}{n}\X_{\hat{S}_j} \Tr\X_{\hat{S}_j} + \lam\W^{-1})^{-1}\tfrac{1}{n}\X_{\hat{S}_j} \Tr \y.
$$

\noindent This suggests the hat matrix 
$$\tilde{\H} = \X_{\hat{S}_j}(\tfrac{1}{n}\X_{\hat{S}_j} \Tr\X_{\hat{S}_j} + \lam\W^{-1})^{-1}\tfrac{1}{n}\X_{\hat{S}_j} \Tr$$
leading to a partial projection matrix of 
$$\tilde{\Q}_{\hat{S}_j} = \I_n - \X_{\hat{S}_j}(\tfrac{1}{n}\X_{\hat{S}_j} \Tr\X_{\hat{S}_j} + \lam\W^{-1})^{-1}\tfrac{1}{n}\X_{\hat{S}_j} \Tr.$$ 
Implementation follows exactly along the same lines as Algorithm~\ref{alg:singlelambda-ci} but with $\tilde{\Q}_{\hat{S}_j}$ instead of $\Q_{\hat{S}_j}$ where in $\tilde{\Q}_{\hat{S}_j}$, $\lam = \lam_{\CV}$. We refer to the intervals constructed using this likelihood as the Local Quadratic Approximation Posterior (LQA-P) intervals.

To build intuition for this adjustment, consider the case of an orthonormal setup. That is, let $\frac{1}{n}\X_{\hat{S}_j} \Tr\X_{\hat{S}_j} = \I_n$, then
$$
\begin{aligned}
\tilde{\Q}_{\hat{S}_j} &= \I_n - \X_{\hat{S}_j}(\I_n + \lam\W^{-1})^{-1}\tfrac{1}{n}\X_{\hat{S}_j} \Tr\\
&= \I_n - \tfrac{1}{n}\X_{\hat{S}_j}\tilde{\W}\X_{\hat{S}_j} \Tr \\
\end{aligned}
$$

\noindent where $\tilde{\W} = \diag\left(|\bh_k| / (|\bh_k| + \lam)\right)_{k \in \hat{S}_j}$. So, when features are orthogonal, the projection weights reflect the proportion of the lasso estimate relative to its pre-shrinkage value ($|\bh_k| + \lam = |z_k|$).

\section{Results}
\label{sec:results}

In Section~\ref{Sec:coverage2}, we benchmark PIPE-P and LQA-P against RL-P in what might be considered the ``ideal'' setting, where the distribution of coefficients matches the Laplace distribution implicit in the lasso penalty. Section~\ref{Sec:Distributions} then explores alternative $\bb$ distributions, and Sections~\ref{Sec:correlation2}~and~\ref{Sec:highcorr} assess how each method responds to correlation among features. The nominal coverage rate is fixed at 80\% in all experiments.

\subsection{Coverage}\label{Sec:coverage2}

\begin{figure}[hbt!]
    \begin{center}
    \includegraphics[width=0.8\linewidth]{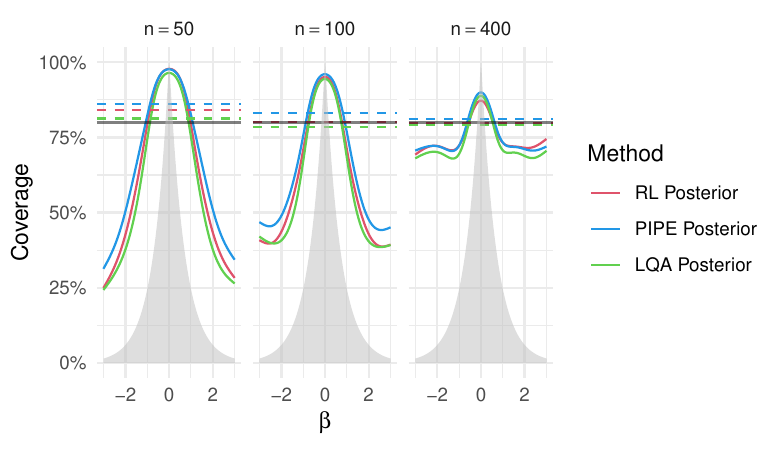}
    \caption{\label{fig:1-man2} Results from the simulation described in Section~\ref{Sec:coverage2}. The fitted curves are from Binomial GAMs fit with coverage being modeled as a smooth function of $\beta$. The dashed lines represent the average for each method across all 1000 independently generated datasets and the solid black line indicates the nominal coverage rate. The shaded distribution in the background depicts the Laplace distribution the $\beta$s were drawn from. }
    \end{center}
\end{figure}

In this section, we generate 1000 independent data sets, each simulated as follows. Each coefficient $\beta_j$ was set to the $j/(p+1)$ quantile of a Laplace distribution. We then scaled the coefficients so that $\bb \Tr\bb = \sigma^2 = 100$. With independent features, this results in a signal-to-noise ratio (SNR) of 1. Then, each element of $\X$ was generated independently from a $N(0, 1)$ with $p = 101$ and three different values of n: 50, 100, and 400. Finally, $\y$ was generated as $\y = \X\bb + \bvep$, where $\veps_i \iid N(0, \sigma^2)$. The results for this simulation are provided in Figure~\ref{fig:1-man2}.

Qualitatively, all three methods display similar coverage profiles. In particular, coverage is high when the true coefficient is near zero and falls as $|\beta|$ grows. This is a direct consequence of the lasso penalty: the lasso penalty shrinks estimates towards zero --- when the true value of the coefficient is near zero, this increases the coverage above the nominal value. As $|\beta|$ increases, more bias is introduced, resulting in diminished coverage. This pattern is less pronounced when $\lam$ is small, as the penalty has less impact --- this is what happens as $n$ gets larger and $\lam_{\CV}$ tends to be smaller. There are, however, slight differences between methods. The average coverage of PIPE-P tends to be slightly more conservative, and while this decreases with increasing sample size, it is the slowest to converge to the nominal coverage rate. Meanwhile, LQA-P has coverage nearly exactly equal to the nominal coverage rate for all values of $n$.

\subsection{Robustness to coefficient distribution}\label{Sec:Distributions}

Although the methods were motivated under the idea that average coverage should hold if the empirical distribution of $\beta$ values resembles the prior implied by the lasso penalty, here we show that they are robust to this assumption.

Table~\ref{tab:1-man2} displays the average coverage for each method across $1000$ simulations under 8 coefficient distributions with $p = 101$ and 4 sample sizes: $n$ = 50, 100, 400, and 1000. As before, to maintain the specified SNR of 1, $\bb$ is normalized. Prior to normalization, Sparse 1 had $\bb_{1-10} = \pm(0.5, 0.5, 0.5, 1, 2)$ with the rest equal to zero, Sparse 2 had $\bb_{1-30}$ as equally spaced quantiles from $N(0, 1)$ with the rest equal to zero, and Sparse 3 had $\bb_{1-50}$ as equally spaced quantiles from $N(0, 1)$ with the rest equal to zero. For the T distribution, the degrees of freedom was set to 3 and for Beta, the distribution was generated from Beta(0.1, 0.1) - 0.5, prior to normalization. $\X$ was generated independently from a $N(0, 1)$ and $\y$ was generated as $\y = \X\bb + \bvep$, where $\veps_i \iid N(0, \sigma^2)$.

\begin{table}[hbt!]
    \centering

\begin{tabular}[t]{>{}cclcccc}
\toprule
\multicolumn{3}{c}{  } & \multicolumn{4}{c}{Sample Size} \\
\cmidrule(l{3pt}r{3pt}){4-7}
  & Distribution & Method & 50 & 100 & 400 & 1000\\
\midrule
\includegraphics[width=0.67in, height=0.17in]{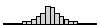} & Laplace & RL Posterior & 84.2\% & 80.1\% & 79.7\% & 80.1\%\\
 &  & PIPE Posterior & 86.1\% & 83.1\% & 81.2\% & 80.4\%\\
 &  & LQA Posterior & 81.3\% & 78.4\% & 79.2\% & 79.8\%\\
\includegraphics[width=0.67in, height=0.17in]{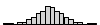} & T & RL Posterior & 83.7\% & 79.4\% & 79.5\% & 80.0\%\\
 &  & PIPE Posterior & 85.7\% & 82.6\% & 80.8\% & 80.1\%\\
 &  & LQA Posterior & 80.8\% & 77.7\% & 78.8\% & 79.6\%\\
\includegraphics[width=0.67in, height=0.17in]{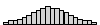} & Normal & RL Posterior & 82.2\% & 76.7\% & 79.4\% & 80.0\%\\
 &  & PIPE Posterior & 84.4\% & 80.3\% & 80.4\% & 80.0\%\\
 &  & LQA Posterior & 78.9\% & 74.9\% & 78.4\% & 79.6\%\\
\includegraphics[width=0.67in, height=0.17in]{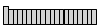} & Uniform & RL Posterior & 80.1\% & 73.3\% & 79.2\% & 80.0\%\\
 &  & PIPE Posterior & 82.9\% & 77.3\% & 79.9\% & 80.0\%\\
 &  & LQA Posterior & 76.5\% & 71.2\% & 78.1\% & 79.7\%\\
\includegraphics[width=0.67in, height=0.17in]{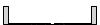} & Beta & RL Posterior & 78.6\% & 68.6\% & 79.1\% & 80.1\%\\
 &  & PIPE Posterior & 81.7\% & 73.8\% & 79.0\% & 80.0\%\\
 &  & LQA Posterior & 74.4\% & 66.3\% & 77.5\% & 79.9\%\\
\includegraphics[width=0.67in, height=0.17in]{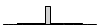} & Sparse3 & RL Posterior & 85.2\% & 82.4\% & 81.7\% & 81.9\%\\
 &  & PIPE Posterior & 86.8\% & 85.0\% & 83.1\% & 82.4\%\\
 &  & LQA Posterior & 82.5\% & 81.2\% & 81.7\% & 81.9\%\\
\includegraphics[width=0.67in, height=0.17in]{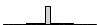} & Sparse2 & RL Posterior & 87.5\% & 85.6\% & 84.4\% & 84.4\%\\
 &  & PIPE Posterior & 88.8\% & 87.6\% & 85.8\% & 85.1\%\\
 &  & LQA Posterior & 85.5\% & 84.7\% & 84.8\% & 84.7\%\\
\includegraphics[width=0.67in, height=0.17in]{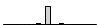} & Sparse1 & RL Posterior & 92.7\% & 91.0\% & 90.1\% & 90.3\%\\
 &  & PIPE Posterior & 93.2\% & 91.9\% & 90.9\% & 90.7\%\\
 &  & LQA Posterior & 91.0\% & 90.5\% & 90.5\% & 90.5\%\\
\bottomrule
\end{tabular}
    \caption{\label{tab:1-man2} Average coverages from the simulation described in Section~\ref{Sec:Distributions}. The nominal coverage rate is 80\%. The first column of the table gives a depiction of the respective distributions.}
\end{table}

Across the three methods, average coverage is broadly comparable. At the smallest sample size ($n = 50$) the differences are most pronounced. PIPE-P is the most conservative and LQA-P is the least. However, as $n$ increases, all three methods approach the nominal 80\% level provided $\bb$ is not generated with a point mass at zero. The coverage pattern is intuitive: distributions that puts more mass / density at zero than the Laplace result in overcoverage, whereas distributions with less mass / density at zero yield undercoverage. Consequently, coverage in highly sparse settings is higher --- in particular, the LQA-P and PIPE-P intervals by construction will always contain $0$ if the variable is not selected. By contrast, when $\bb$ matches a Beta distribution unfavorable for the lasso, the coverage is lowest, dropping to around 70\% when $n = 100$ before rising back to nearly the nominal level of 80\% for large sample sizes.

\subsection{Correlation and Average Coverage}
\label{Sec:correlation2}

We next examine the effect of feature correlation, which is precisely the setting where the three procedures diverge most noticeably. The key distinction lies in how they handle the variance projection: PIPE-P and RL-P rely on full projections, whereas LQA-P introduces a partial projection. This difference is especially relevant when features are correlated. With strong correlations, the projection of $\x_j$ onto $\cC^\bot(\X_{\hat{S}_j})$ shrinks substantially; a full projection discards all of the non-independent signal in feature $j$. In contrast, a partial projection removes only a proportion of the shared information, reflecting the fact that selected features are shrunk rather than fully active.

\begin{figure}[hbt!]
    \begin{center}
    \includegraphics[width=\linewidth]{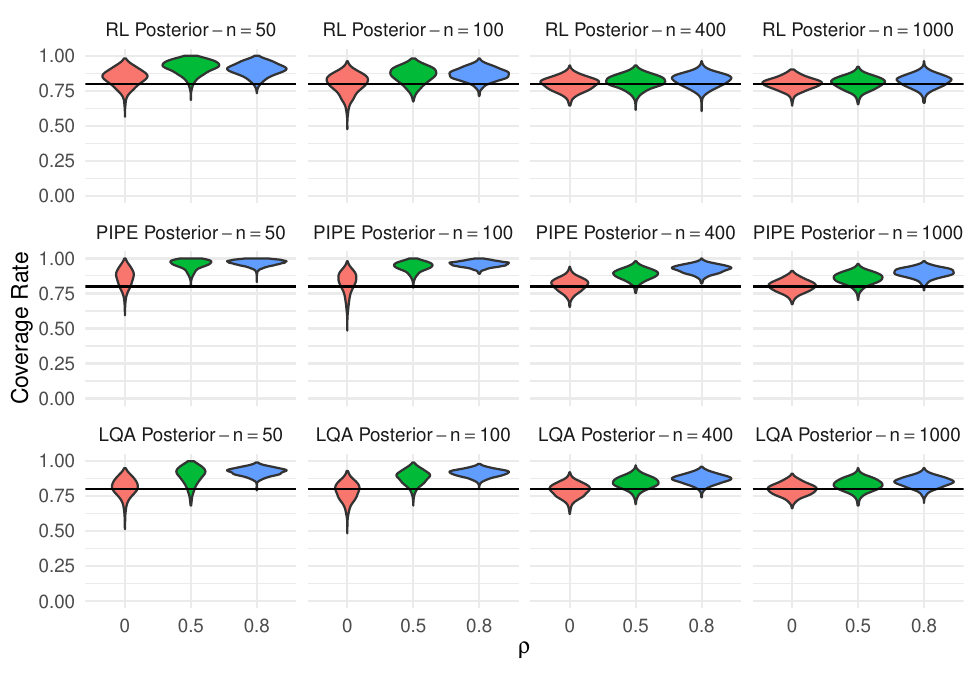}
    \caption{\label{fig:2-man2} Presents results for the simulation described in Section~\ref{Sec:correlation2}. The violin plots are the distributions of average coverages across 1000 simulated datasets for the three methods and across three different levels of autoregressive correlation among the covariates, $\rho = 0 \text{ (no correlation)}, 0.5, 0.8$. For this simulation, p = 101 with $\bb$ set to be equally spaced quantiles from a Laplace distribution. The results for each level of correlation are presented across four different sample sizes, n = p/2, p, 4p, 10p. The horizontal black line provides reference for the 80\% nominal coverage rate.}
    \end{center}
\end{figure}

Other than the addition of an autoregressive correlation structure in which $\cor(\x_i, \x_j) = \rho^{|i-j|}$, the set up of the simulation for the results displayed in Figure~\ref{fig:2-man2} is the same as for the simulation described in Section~\ref{Sec:coverage2}. The violin plots provide the distributions of average coverages across 1000 simulated datasets for four values of $n$ and three values of $\rho$. For each $n$, the amount of correlation is increased from $\rho = 0$ to $0.5$ to $0.8$. The results for RL-P are in the top row, PIPE-P in the middle row, and LQA-P in the bottom row.

PIPE-P is the most conservative in the presence of feature correlation, and average coverage remains above nominal even when $n$ is large. The coverage of LQA-P intervals, on the other hand, lies in between PIPE-P and RL-P: still conservative in the presence of correlation, but markedly less so than PIPE-P. 

\begin{figure}[hbt!]
    \begin{center}
    \includegraphics[width=0.8\linewidth]{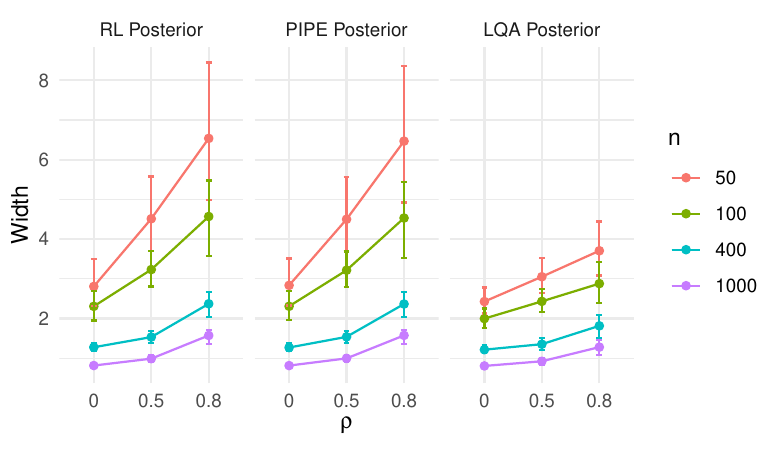}
    \caption{\label{fig:3-man2} Presents results for the simulation described in Section~\ref{Sec:correlation2}. The plots give the median and Q1 and Q3 widths of the intervals across 1000 simulated datasets for the three methods, varying both the level of autoregressive correlation among the covariates, $\rho = 0 \text{ (no correlation)}, 0.5, 0.8$ and the sample size, n = p/2, p, 4p, 10p. For this simulation, p = 101 and $\bb$ was set to be equally spaced quantiles from a Laplace distribution.}
    \end{center}
\end{figure}

Figure~\ref{fig:3-man2} is based on the same simulation results as Figure~\ref{fig:2-man2} and displays the median and inter-quartile range (Q1, Q3) of the interval widths across the 1000 replicates. PIPE-P and RL-P behave similarly; as feature correlation rises, the intervals get wider, with the effect most pronounced at smaller sample sizes. The widths of the LQA-P intervals also increase with correlation, however, the rise is much less sharp and intervals are considerably narrower on average. Thus, although all three methods attain comparable coverage, LQA-P achieves it with substantially shorter intervals, especially in highly correlated settings. The impact of this behavior has notable implications when the methods are applied to high dimensional data sets, a point that will be returned to in Section~\ref{Sec:Scheetz2006-2}.

\subsection{Correlation and Individual Intervals}
\label{Sec:highcorr}

While the previous section showed that the interval construction methods are robust for average coverage, this hides the impact of correlation on individual interval behavior. In this section, we show the effect of correlation between features on the intervals themselves.

\begin{figure}[hbt!]
    \begin{center}
    \includegraphics[width=\linewidth]{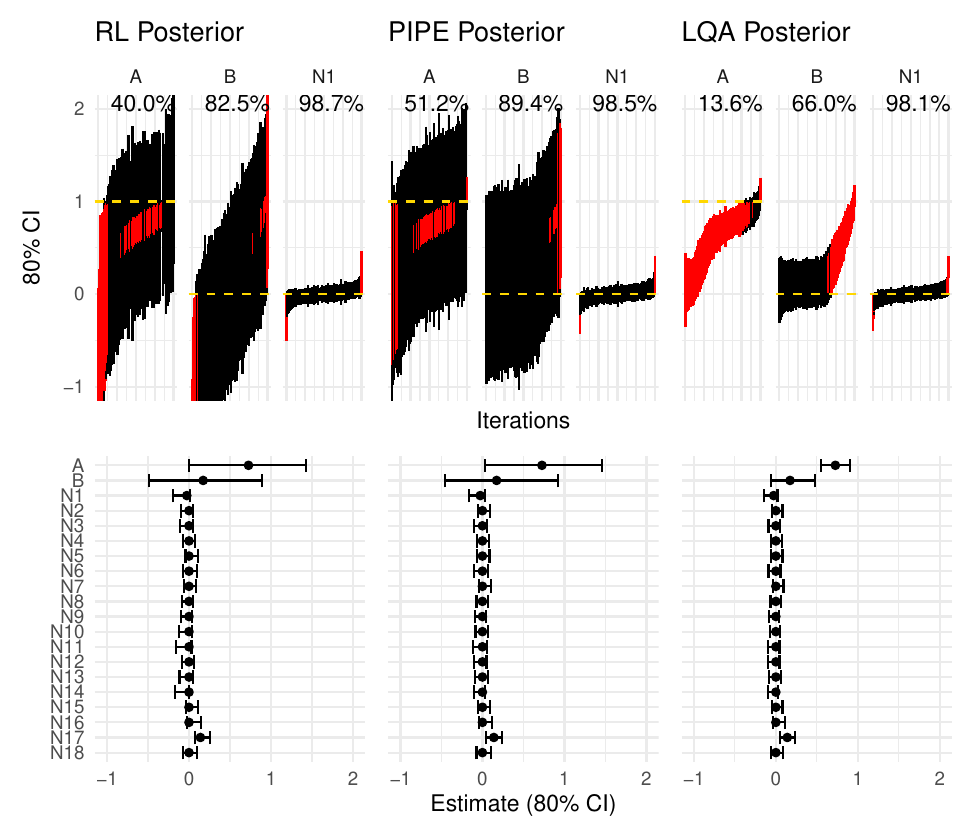}
    \caption{\label{fig:4-man2} Provides results for simulation described in Section~\ref{Sec:highcorr}. The bottom plots show a single example of the intervals produced by the RL Posterior (left), the PIPE Posterior (middle), and the LQA Posterior (right) methods for the first 20 variables from one (randomly chosen) of the datasets where both feature $A$ and $B$ were selected. The top plots summarize the HDIs for the variables $A$, $B$, and $N1$ across the 1000 simulations. All 1000 HDIs are plotted, sorted by their midpoint, with those colored red that did not contain the true coefficient value (indicated by the horizontal dashed gold line).}
    \end{center}
\end{figure}

In this example, $n = p = 100$. However, only one $\beta_j$ is non-zero: $\beta_{A} = 1$ and $\beta_{B}, \beta_{N1}, \ldots, \beta_{N98} = 0$. Additionally, the data are simulated such that $\cor(\x_{A}, \x_{B}) = .99$ but all of the N (noise) variables are uncorrelated with $A$, $B$, and each other. The distribution of $\X$ and $\y$ is unchanged from Section~\ref{Sec:coverage2}, although here $\sigma^2 = 1$.
Figure~\ref{fig:4-man2} depicts the results from $1000$ simulated data sets. The top shows 1000 HDIs for for 3 features: $A$, $B$, and $N1$; the HDIs are colored black if they contain the true parameter value and red if they do not. The bottom displays intervals for the first 20 variables for a randomly chosen example data set where both feature $A$ and $B$ were selected. Although $A$ is the only feature with a true signal, its high correlation with $B$ produces uncertainty about which feature contains the signal, or whether both $A$ and $B$ contain signal. As we will see, PIPE-P handles this uncertainty the same as RL-P while LQA-P differs considerably in its treatment of the uncertainty. 

For RL-P and PIPE-P, when A is selected, the interval for B is wide and vice versa. This leads to intervals which are highly polarized in their widths --- either narrow or very wide. While the RL-P and PIPE-P intervals have the same width, the PIPE-P intervals are centered on the lasso estimates and, as a result, are closer to zero.

The LQA intervals for features A and B are strikingly different from the RL-P and PIPE-P intervals. The intervals are much narrower and do not exhibit the bimodal narrow/wide behavior of the other two methods. This occurs because LQA-P uses a partial projection for the variance, in which the width changes gradually as a coefficient is introduced into the model, instead of fully projecting onto the feature as soon as it is selected. This results in lower coverage, although nominal average coverage is maintained.

Clearly, the full and partial projection approaches produce very different intervals. It is less clear which intervals are ``correct.'' Looking at coverage is not sufficient to answer this: any method can achieve nominal coverage by producing pathological intervals that alternate between being too wide and too narrow. For an inferential method to be useful, its intervals must also provide information that is relevant and interpretable in the specific context, conveying what is known and what is uncertain about the parameter at hand.

Consider the case of two highly correlated features selected into the model, as shown in the bottom row of Figure~\ref{fig:4-man2}. RL-P and PIPE-P use variance based on a full projection, which --- analogous to the variance formula in ordinary least squares --- treats the correlation as shared uncertainty, producing wide and similar width intervals for both features. In contrast, LQA-P uses variance based on a partial projection, which plays a role similar to the stabilizing effect of penalization in Ridge regression. Although the intervals for A and B remain wider than those for null features, they are substantially narrower than their RL-P and PIPE-P counterparts. In addition, the LQA-P interval for A is narrower than that for B. Unlike the full projection approaches, the LQA-P widths depend not only correlation but also the estimates themselves. Since $\bh_B$ is mostly shrunken to zero, it introduces less uncertainty for feature A than the other way around.

\section{An argument for biased intervals}
\label{sec:biased-intervals}

The intervals developed in this manuscript are biased and designed to achieve nominal average coverage, which by definition allows coverage to vary across coefficients. A practitioner focused only on the selected variables might view this unevenness as problematic: perhaps average coverage is preserved only because null coefficients enjoy high coverage while the selected coefficients --- the ones of substantive interest --- have much lower coverage. From this perspective of ``relevant'' average coverage, defined as the average coverage among the selected variables, biased intervals might seem unsatisfactory, since the greatest bias typically occurs for the largest effects. To address this concern, we examine whether the proposed PIPE-P and LQA-P intervals provide adequate relevant average coverage compared to commonly used debiased methods including Selective Inference, the De-sparsified Lasso, and the Relaxed Lasso.

The De-sparsified Lasso \citep{ZhangZhang2014}, as the name suggests, modifies the original point estimates from the lasso in such a way that they are no longer sparse or shrunk towards zero, thereby facilitating classical approaches to inference. Alternatively, Selective Inference \citep{Lee2016,Tibshirani2016} aims to account for the uncertainty in model selection by conditioning on the selected model. This conditioning is also a form of bias correction, albeit indirectly. Note that through conditioning on the selected model, Selective Inference only provides intervals for the covariates that were selected. Lastly, the Relaxed Lasso is a is a two-step procedure which uses the lasso for variable selection, then refits using only the selected variables with a reduced penalty (commonly no penalty) \citep{Meinshausen2007}.

For this simulation, 4 different distributions were considered for $\bb$:
\begin{enumerate}
  \item $p = 16$: $4$ large signals ($\beta = \pm 5$)
  \item $p = 100$: $4$ large signals ($\beta = \pm 5$)
  \item $p = 50$: $10$ moderate signals ($\beta = \pm 1$)
  \item $p = 100$: $50$ small signals ($\beta = \pm 0.5$)
\end{enumerate}

\noindent In all 4 scenarios, $n = 50$, $\X \iid \Norm(0,1)$, and $\y = \X\bb + \epsilon$, with $\epsilon \iid \Norm(0, 1)$. In each scenario, 1000 independent replications were generated. For each data set, $\lambda$ was selected using cross validation, then each method was applied at the selected value of $\lambda$ to construct intervals. In this simulation, unlike in the rest of the manuscript, we calculate relevant average coverage. The coverage and intervals widths for the selected features is summarized in Table~\ref{tab:2-man2}.

\begin{table}[hbt!]
  \centering

\begin{tabular}[t]{lcc}
\toprule
Method & Coverage (mean, sd) & Width (median, Q1--Q3)\\
\midrule
\addlinespace[0.3em]
\multicolumn{3}{l}{\textbf{Scenario 1}}\\
\hspace{1em}Desparsified Lasso & 61.3\% (17.6\%) & 0.366 (0.336, 0.397)\\
\hspace{1em}Selective Inference & 78.1\% (20.9\%) & 0.780 (0.536, 1.417)\\
\hspace{1em}Relaxed Lasso & 62.6\% (17.9\%) & 0.381 (0.353, 0.412)\\
\hspace{1em}PIPE Posterior & 69.9\% (17.4\%) & 0.370 (0.340, 0.401)\\
\hspace{1em}LQA Posterior & 68.6\% (17.5\%) & 0.358 (0.330, 0.391)\\
\addlinespace[0.3em]
\multicolumn{3}{l}{\textbf{Scenario 2}}\\
\hspace{1em}Desparsified Lasso & 38.8\% (19.0\%) & 0.435 (0.408, 0.466)\\
\hspace{1em}Selective Inference & 62.7\% (26.8\%) & 103.603 (52.528, Inf)\\
\hspace{1em}Relaxed Lasso & 44.2\% (19.6\%) & 0.346 (0.318, 0.376)\\
\hspace{1em}PIPE Posterior & 52.2\% (18.6\%) & 0.412 (0.384, 0.442)\\
\hspace{1em}LQA Posterior & 49.5\% (17.9\%) & 0.391 (0.363, 0.425)\\
\addlinespace[0.3em]
\multicolumn{3}{l}{\textbf{Scenario 3}}\\
\hspace{1em}Desparsified Lasso & 55.5\% (13.7\%) & 0.380 (0.341, 0.421)\\
\hspace{1em}Selective Inference & 76.1\% (27.3\%) & 33.472 (19.212, 101.228)\\
\hspace{1em}Relaxed Lasso & 53.6\% (15.3\%) & 0.395 (0.344, 0.444)\\
\hspace{1em}PIPE Posterior & 74.6\% (11.7\%) & 0.471 (0.419, 0.523)\\
\hspace{1em}LQA Posterior & 67.1\% (12.0\%) & 0.400 (0.362, 0.442)\\
\addlinespace[0.3em]
\multicolumn{3}{l}{\textbf{Scenario 4}}\\
\hspace{1em}Desparsified Lasso & 51.3\% (19.7\%) & 0.744 (0.518, 0.971)\\
\hspace{1em}Selective Inference & 68.4\% (35.8\%) & 48.501 (14.942, Inf)\\
\hspace{1em}Relaxed Lasso & 50.3\% (20.2\%) & 0.646 (0.464, 0.830)\\
\hspace{1em}PIPE Posterior & 87.4\% (12.3\%) & 0.921 (0.816, 1.024)\\
\hspace{1em}LQA Posterior & 74.9\% (18.6\%) & 0.707 (0.580, 0.823)\\
\bottomrule
\end{tabular}
  \caption{\label{tab:2-man2} Relevant average coverage (average coverage for features with non-zero lasso estimates) and median widths for two biased interval construction methods (PIPE-P and LQA-P) compared to three debiased methods across 4 set ups described in Section~\ref{sec:biased-intervals}.}
\end{table}

In Scenarios 1-3, while the coverage for the selected covariates is less than the nominal coverage rate for all methods, the two biased methods (PIPE-P and LQA-P) are only outperformed in terms of coverage by Selective Inference. In Scenario 4, PIPE-P and LQA-P provide coverage closer to the nominal rate than all three debiased methods. Although Selective Inference has better performance in Scenarios 1-3, it is important to note that Selective Inference produces very wide intervals --- in the first scenario, Selective Inference produces intervals over twice as wide as the other methods and orders of magnitude wider in the other scenarios. Aside from Selective Inference, the other methods all tend to produce intervals of similar width.

While these four scenarios are obviously not exhaustive, they suggest that the argument against biased intervals on the grounds of ``relevant coverage'' is more complex than anticipated. These results suggest that even if one is only interested in intervals for the selected features, the biased methods proposed here may perform better than debiased alternatives, and that attempting to reverse the bias imposed by the lasso penalty may cause more harm than good.

\section{Study of gene expression in the mammalian eye}
\label{Sec:Scheetz2006-2}

\begin{figure}[hbt!]
    \begin{center}
    \includegraphics[width=0.9\linewidth]{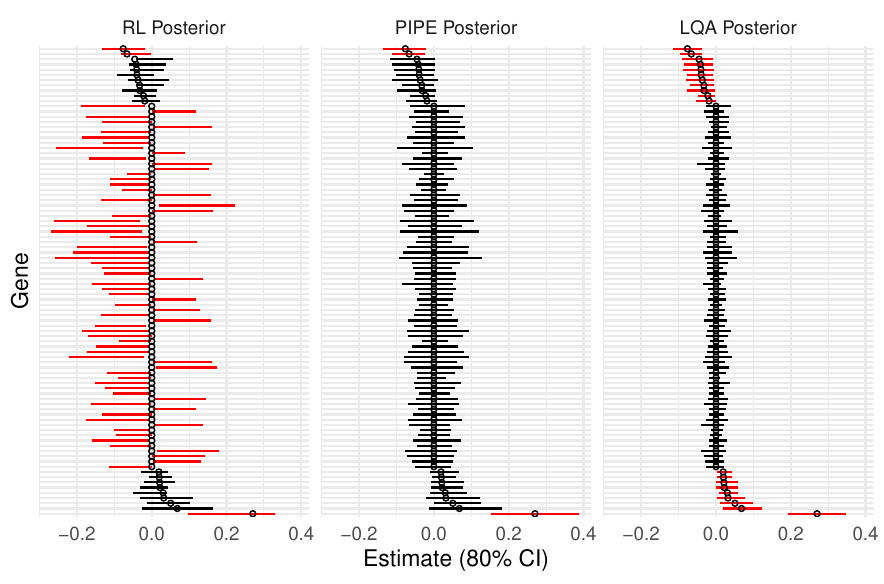}
    \caption{\label{fig:5-man2} Intervals from three methods for the 90 genes in the Scheetz2006 dataset (Section~\ref{Sec:Scheetz2006-2}) for which at least one method yielded an interval that excluded zero. Intervals are colored red if they do not contain zero. Gene labels are excluded for compactness.}
    \end{center}
\end{figure}

In this section, we analyze data from a gene expression study in mammalian eyes. \citet{Scheetz2006} measured the RNA levels from the eyes of 120 rats. A total of 18976 probes exhibited expression in at least a subset of samples. For this analysis we treat one of the genes, Trim32, as the outcome since it is known to be linked to the genetic disorder Bardet-Biedl Syndrome (BBS). The remaining 18975 genes were used as predictors with the goal of determining other genes whose expression is associated with Trim32 and thus may also contribute to BBS. In Figure~\ref{fig:5-man2}, intervals are displayed from three methods for the 90 genes for which at least one method yielded an interval that excluded zero.

This analysis reveals that RL-P intervals can behave in undesirable ways in high-dimensional problems. Recall that RL-P intervals, unlike PIPE-P and LQA-P intervals, are not guaranteed to contain the lasso point estimate. In particular, Figure~\ref{fig:5-man2} reveals that for 70 genes, the RL-P interval does not contain 0 despite the fact that the lasso point estimate is zero. By construction, this cannot occur for PIPE-P or LQA-P.

Comparing PIPE-P and LQA-P, the LQA-P intervals are considerably narrower, consistent with the simulation results. In particular, while both intervals are centered at the same points (the lasso estimates), 20 of the LQA-P intervals exclude zero compared to only 3 of the PIPE-P intervals. While it is difficult to say which approach is ``correct'', the PIPE-P intervals are based on a full projection onto the column space of all 66 genes selected by the lasso, which accounts for the wide intervals given that $n$ is only 120.

\section{Discussion}

Our motivation in constructing high-dimensional intervals is to produce a single, coherent set of coefficient estimates, predictions, and intervals that are all consistent with each other. In our earlier paper \citep{AvgCov}, we introduced the idea of average coverage, arguing that high-dimensional intervals would be more consistent with lasso estimates if we dropped the requirement that every interval had to maintain correct coverage --- this can never be consistent with regularized estimation --- and replaced it with a requirement that the average coverage of all the intervals is maintained at a prespecified rate. 

Although the Relaxed Lasso Posterior approach proposed in that manuscript improved alignment between high-dimensional intervals and corresponding point estimates, the intervals were not completely satisfying in the sense that they were not guaranteed to contain the lasso point estimates. To address this, we have now introduced two methods, PIPE-P and LQA-P, for constructing high-dimensional intervals that are centered on the lasso estimates. The two methods condition on the features selected by the lasso in different ways, and as a result the widths of the intervals can differ substantially. Nevertheless, both approaches provide meaningful ways to represent our uncertainty about the penalized estimates and bring us closer to the goal of a coherent inferential framework.

\newpage

\bibliographystyle{ims-nourl}

\end{document}